\documentclass[12pt]{iopart}

\newcommand{\myskip}[1]{}

\renewcommand{\d}{{\rm d}}

\newcommand{\BEQ}{\begin{eqnarray}}
\newcommand{\EEQ}{\end{eqnarray}}
\newcommand{\BEA}{\begin{eqnarray}}
\newcommand{\EEA}{\end{eqnarray}}
\newcommand{\nn}{\nonumber }
\renewcommand{\d}{{\rm d}}

\newcommand{\eff}{{\rm eff}}

\usepackage{graphicx}
\usepackage{dcolumn}
\usepackage{bm}

\newcommand{\mBDs}{$\mu$BDs}

\begin{document}

\title[ The Mass Function of Primordial Rogue Planet MACHOs in quasar nanolensing]
{ The Mass Function of Primordial Rogue Planet MACHOs in quasar nanolensing} 
 
 \author{
Rudolph E. Schild$^{1}$,
Theo M. Nieuwenhuizen$^{2}$ and Carl H. Gibson$^{3}$}

\address{$^1$ Harvard-Smithsonian Center for Astrophysics, 60 Garden Street, Cambridge, MA02138,
USA\\
$^{2}$Institute for Theoretical Physics, 
Science Park 904, P.O. Box 94485, Amsterdam, the Netherlands\\
$^3$
Mech. and Aerospace. Eng. \& Scripps Inst. of Oceanography. Depts., UCSD, La Jolla, CA
92093, USA }

\ead{custserv@iop.org}

\begin{abstract}
The recent Sumi et al (2010, 2011) detection of free roaming planet mass
MACHOs in cosmologically significant numbers recalls their original
detection in quasar microlening studies (Schild 1996, Colley and Schild
2003).
We consider the microlensing signature of such a population, and find that
the nano-lensing (microlensing) would be well characterized by a statistical
microlensing theory published previously by Refsdal and Stabel (1991).
Comparison of the observed First Lens microlensing amplitudes with the 
theoretical prediction
gives close agreement and a methodology for determining the slope of the
mass function describing the population. Our provisional estimate of the power
law exponent in an exponential approximation to this distribution is
$2.98^{+1.0}_{-0.5}.$ where a Salpeter slope is 2.35.

\end{abstract}


\maketitle





\section{Introduction}

The recent detection of planetary mass MACHOS seen in high-cadence searches
toward the Galactic Center and the Large Magellanic Cloud (LMC) (Sumi et al, 2010, 2011)
suggest that a significant population of dark planet-mass MACHOs populate
the halo of our Galaxy,  which may constitute the Galactic dark matter.
This would (partly) explain the ``missing baryon problem'', 
the fact that about 90\% of the baryons in the solar neighborhood are unaccounted for.
Indeed, below we clarify this from similar detections made in last decades with 
micro-lensing and even nano-lensing in the Q0957+561
A,B gravitational lens system (the so-called  First Lens). 
Insofar as the observed MACHOs are planetary mass bodies dominated by
hydrogen and 26\% in weight of helium, other modes of their detection as extreme scattering events
seen in the line of sight to ordinary quasars, and 
as the  sources of the ubiquitous ``dust'' emission at temperatures of minimally 15 K
and of the ``mysterious radio events'', may also be signaling their detection (Nieuwenhuizen et al 2010).

In the following we re-consider the expected microlensing signature,
optical depths, and mass spectrum of the observed planet mass MACHO population.
Presently there is confusion in the literature about what to call these
objects, because there is increasing awareness from emerging statistics of
ubiquity of orbiting planets accompanying ordinary stars that the many
orbital interactions should frequently result in escape from the initial
orbiting planetary systems. To avoid confusion, we call free-roaming
planetary mass condensed objects ``escapees'' if originally
formed in pre-stellar accretion discs, and ``micro brown dwarfs'' (\mBDs) 
  if formed primordially by gas fragmentation. 
Of course we still call ordinary star-orbiting bodies planets.

\subsection{Quasar Microlensing and the missing baryons problem}

The existence of a population of planet mass microlensing objects, also 
called MACHOs, was first inferred from quasar microlensing studies by
Schild (1996) when it was discovered that no value close to the accepted 
time delay would remove the pattern of daily sampled quasar brightness
fluctuations. The subject was complicated by the fact that to securely 
recognize the signal, it would be necessary to measure the double image
quasar time delay to a fraction of a day. This was accomplished by an 
international around-the-world consortium (Colley et al, 2002, 2003) 
whose 417.1-day time delay value
still stands as the most precise time delay ever measured. 
Re-analysis of an earlier observation of 5 consecutive 
nights of continuous brightness monitoring data produced a microlensing
event of duration 5 hours (quasar proper time), by Colley and Schild, (2003).

Such rapid events can realistically be understood only as resulting from 
microlensing. Recall that a quasar is approximately 100 times brighter than
our Milky Way galaxy. If attributed to the quasar, the  observed 1 
percent brightness fluctuation in
the quasar light is energetically equivalent to the entire Milky Way
luminosity being switched on and off in only 5 hours -- which is an implausible scenario.
\footnote{There is a counter indication for this argument. Colley and Schild (2003) report a 5 night observation
for both the A and the B image in periods corresponding to a common quasar time.
In each day (night) of the 5 day observation by Colley and Schild, there is a common
fluctuation in both  images of the quasar with period of
approximately one day.  On the last night the trends in both images are different, which is interpreted as an additional lensing event on
 top of a common trend.  However, the 24/7 monitoring of the quasar for 10 days in 2001 did not exhibit this
trend (Colley et al, 2003). This conundrum has to be resolved with new observations.}
For observation from the earth,  the intensity fluctuation can be accomplished by the gravitational field 
of a microlensing compact object. Indeed, an
observed negative brightness change is in fact a re-direction of the 1\% part 
of the enormous
luminosity away from the line of sight of the affected quasar image. 
Likewise, a positive 1\% change 
is due to a lensing effect that focuses more light in the observer's direction.

The original rapid microlensing detection was greeted with strong interest,
because it was immediately recognized that it could not be caused by stars
(Catalano 1997). Since it was obviously seen as a continuous brightness fluctuation
pattern at approximately unit optical depth, it must represent the
detection of the dark baryonic matter.
But the interest faded away when the EROS and MACHO
consortia did not observe similar MACHOs in front of the Magellanic Clouds.
However, the quasar data themselves have never been questioned and related effects
were observed on other, lensed quasars, as commonly discussed in the context of
measurement of time delay (Burud et al, 2000, 2002; Paraficz et al, 2006, 
Vakulik et al, 2007, 2008).
To settle this dispute, we intend to redo a search in front of the Large Magellanic Cloud
(Schild et al, 2012).

 To explain the quasar observations, other possibilities than MACHOs were 
considered as well,
such as hypothetical orbiting luminous blobs in the accretion disc, for which, however,
there had never previously been evidence (Gould and Miralda-Escude, 1996).
Finally, a series of simulations of orbiting luminous blobs and obscuring
clouds by Wyithe and Loeb
(2002) produced brightness curves that could be compared to observations.
Simulations for orbiting dark spots and bright spots microlensed by
patterns of cusps originating in the lens galaxy show that the longer
duration events have smaller brightness amplitudes than shorter duration
events (Figure 6).
But the wavelet analysis of the observed microlensing
brightness fluctuations by Schild (1999) consistently showed an opposite  pattern
of larger brightness fluctuation amplitude for longer duration events.
Moreover, in general the simulated brightness curves do not look
like the observational results, and in particular do not show the observed
feature of equal positive and negative events (Schild 1999).

An additional simulation
showed that for the process championed by Schild (1996), with quasar
structure microlensed by cusps originating in the lens galaxy, larger
amplitude brightness effects were always found for longer duration events
(Wyithe and Loeb, 2002, Fig. 9). The wavelet analysis of the microlensing
brightness fluctuations by Schild (1999) consistently showed this pattern
of larger brightness amplitude for longer duration events.

Since the Wyithe and Loeb (2002) simulation only covered stellar to  Jupiter microlensing
masses and not to the Earth masses or smaller, as implied by the Colley and Schild (2003)
event, the targeted simulation by Schild and Vakulik (2003) for The First
Lens, Q0957, is of more relevance. It demonstrates microlensing events 
caused not by orbiting blobs, but rather by a drift pattern of microlensing
cusps originating in the lens galaxy and microlensing the discrete luminous
quasar structures, creating a continuous pattern of brightness cusps at the 
observed $1\%$ level with event durations of a few days, caused by
microlens masses $10^{-5}M_\odot$.

Thus in this paper we return to the interpretation of the quasar microlensing effect as
due to a population of MACHOs.
In the following sections we will first consider in section 2 the statistics describing
the probability of microlensing in the quasar lens system, and then in section 3 we
show how the theory of microlensing for large luminous sources gives an
excellent fit to the observed amplitude of brightness fluctuations. In
section 4 we demonstrate a technique for
determination of the mass function describing the observed masses of
the rogue planet MACHOs. We summarize our conclusions in section 5.

\section {Quasar structure and microlensing observations}

\subsection{On size scales of luminous quasar structure inferred from
  microlensing} 

Technically speaking, the subject of lensing by individual objects within
a resolved object like a distribution of stars in a lens galaxy foreground
to a distant quasar is discussed under the topic of microlensing, although
if the grainey distribution is composed of planet mass objects, it would
more correctly be described as nano-lensing. We follow standard usage and
simply adopt the word microlensing in this report.

It is also important to understand that the microlensing by planet mass
objects resolves the structure of the central regions of quasars, taken to
be black hole or MECO objects (Schild, Leiter, and Robertson 2006; SLR06). 
For one such object, the doubly imaged quasar  Q0957+561 the distance observed between the A and B images is 6.26 arcsec.
It has redshift of source $z_S = 1.43$ and of the lens $z_L = 0.355$, 
the BH mass is $3.6\,10^9 M_\odot$ (SLR06) and the gravitational radius $R_G=GMc^{-2}$ is 
$5.3\cdot 10^{14}$ cm.

We describe the luminour ring at the inner edge of the accretion disc as a torus having two radii of importance; 
the outer radius, which refers to the quasar central radius to the inner edge of the accretion disc, 
and the sectional radius, which is half the thickness of the accretion disc and
therefore the cross sectional radius. Then the outer radius is $R_{\rm outer}=74\,R_G=3.94\,10^{16}$ cm
and the sectional radius is $R_\perp'=1.52 \cdot10^{13}$ cm.
The main part of the quasar light comes from a structure of this radius.
\footnote{\label{fn6} Indeed it is known from the examination of the quasar spectrum that 
at this wavelength  of brightness monitoring, the inner edge of the accretion disk
contributes only $\frac{1}{4}$ of the observed brightness. What is microlensed is a bump in the spectrum, which is the thermal peak of light
originating at the inner accretion disk. It is known that the brightness of this structure is $\frac{1}{4}$ of the total brightness at the monitoring 
wavelength. So if the $\frac{3}{4}$ of brightness coming from the outer larger structure were absent, the quasar would be fainter
overall, and the observed amount from the microlensed structure would be the same. But the microlensing comes from $\frac{1}{4}$
of the quasar brightness. Since we observe fluctuations in the total quasar brightness, the observed
fluctuations should be multiplied by $4$ to get the relative brightness fluctuations that 
 would have been observed if the outer luminous structure were absent.}
We adopt the standard cosmology with $H_0 = 70$ km/s Mpc, $\Omega_c =0.23$,
 $\Omega_B=0.05$, so that $\Omega_M=0.28$, and $\Omega_{\Lambda} = 0.72$.
 The general formula for the angular distance $d_A(z)=H_0^{-1}(1+z)^{-1}\int^{1+z}_1\d x \,x^{-1}(\Omega_L+\Omega_Mx^{3})^{-1/2}$
 yields for the angular distance to the source $d_S\equiv d_A(z_S)=1160$ Mpc and to the lens $d_L\equiv d_A(z_L)=890$ Mpc,
 while the angular distance between them, for light that we observe now, is $d_{SL}=d_S-(1+z_L)(1+z_S)^{-1}d_L= 670$ Mpc.
 The angular radius of the source is  thus $\theta_{\rm outer}=R_{\rm outer}/d_S= 2.14\,\mu{\rm as}$.

The Colley and Schild (2003) observation of an event of duration 12
hours (observer's clock) as an event with an approximately Gaussian profile
instead of a more sharply peaked brightness profile (as from a point
source) 
can be used to give an
approximate dimension to the luminous quasar structure. Assuming as in
SLR06 that the brightness profile results from a microlensing rogue planet 
moving past
the luminous inner edge of its accretion disc, for a standard
cosmological transverse velocity of 600 km/sec, and a quasar distance 
of 1.31 times the lens distance (in the above mentioned cosmology) 
we conclude that the ring-shaped luminous inner edge of the accretion disc
has a thickness dimension approximately equal to the Einstein ring diameter
of the microlens, or  $d'_\perp=1.52 \cdot 10^{13}$ cm for the Colley and Schild (2003) event in the cosmology $H_0=75$ km s$^{-1}$ Mpc$^{-1}$,
$\Omega_M=1$, $\Omega_\Lambda=0$ (as indicated by the prime on $R_\perp'$). 
Hence its angular diameter from that cosmology is $\theta_\perp=d'_\perp/d_A'(z_L)=1.41$ nas,
which in our cosmology corresponds to a radius $d_\perp=1.88\,10^{13}$ cm. The black-hole-centric diameter of the inner 
luminous edge of the accretion disc has been determined
from reverberation to be $7.8 \cdot 10^{16}$ cm (SLR06; Schild and Leiter 2009),
so multiplying by $2\pi$ times $R_\perp$
determines the area of the luminous structure as
 $2.83\cdot10^{30}$ cm$^2$.  In the theory of RS91 the number of
deflectors is the product of their normalized surface density $\sigma$ and 
the surface area,
so we may replace the luminous ring by an equivalent round structure having 
this area for an effective radius of $R_\eff=0.949\,10^{15}$ cm
and effective angle $\theta_\eff=R_\eff/d_L=71.4$ nas.

It is necessary to calculate this luminous area radius to compare with the
radius of the Einstein ring of a microlensing particle. For the Colley and
Schild (2003) microlensing event, the radius of the Einstein ring was
previously calculated from the event duration to be  
$\frac{1}{2} \cdot 1.52 \cdot 10^{13}$ cm $=0.76 \cdot 10^{13}$ cm and therefore
the RS91 requirement that the luminous structure's radius should be at least 5 times
greater than the microlens's Einstein radius is well satisfied. Therefore we may
use the RS91 statistical theory results to describe the relationship of
brightness fluctuation amplitude to microlensing optical depth.

This allows us to use two statistical results that
relate the measured brightness fluctuation amplitude to the area of the
lensing luminous source and to the mass function of the nano-lensing
objects.   

\section{Estimation of the Einstein ring radius and microlensing event duration}

The discussion until now pre-supposes that all of the rogue planet
MACHOs constituting the baryonic dark matter have the same mass. That mass
was predicted to be typically $10^{-7} M_\odot=0.03M_\oplus$ by Gibson (1996), 
who also predicted that
the initial fragmentation would have been immediately followed by an
accretional cascade to larger masses, especially in the times immediately
following fragmentation after recombination, because the MACHOs were hotter
and larger from their gravitational energy release, and because the density
of the universe has been monotonically decreasing since the formation
epoch, increasing the spacings. A more recent estimate of the mass comes out larger,
$3.9\ 10^{-5}M_\odot=13M_\oplus$ (Nieuwenhuizen et al, 2009).

Thus the mass function $F(M)$ describing the number of MACHOs 
by $F(M)\d M$
as a function
of mass would have been modified since the original formation epoch by the
process of accumulation. The prescient statistical microlensing theory of 
Refsdal and Stabell (1991)
that describes the determination of brightness fluctuation amplitudes 
in terms of the Einstein radius of the MACHO and the optical depth, also
demonstrates a formalism for determining the slope the mass function $F(M)$.

This occurs because relative to a simple mass function  with a delta
function for some mass $M$, a mass function with additional masses larger
than $M$ has relatively more large-amplitude events. We shall consider the case of $F(M)$ expressed as a power law, 

\BEQ
\label{FMdM}
F(M) \d M=A\left(\frac{M_\oplus}{M}\right)^{\alpha}\frac{\d M}{M_\oplus},\qquad (M_1<M<M_2) \EEQ
while $F=0$ for $M<M_1$ and for $M>M_2$.
In practice, the mean amplitude of brightness fluctuations is measured from
brightness monitoring, the Einstein Ring diameter is estimated from the
duration of the microlensing events, the diameter of the source is known or
estimated, and the optical depth is known from the overall macro-lensing
model.

We adopt a standard transverse velocity of $v=600$ km/s, based upon the presumption that extreme cosmological departures
from the co-moving expansion velocity are top-limited by the local Great Attractor at approximately 1000 km/s,
and that other velocities, (e.g., Earth orbit, Galaxy rotation, Galaxy cluster motion, etc) are of order 50-250 kms/s
and uncorrelated. So a value considered accurate to a factor $\sqrt{2}$ uncertainty, is ordinarily taken to be 600 km/s.

The opening angle of the Einstein Ring  is 

\BEQ \label{thetaE=}
 \theta_E(M)=\sqrt{\frac{4GM}{c^2}\ \frac{d_{SL}}{d_S d_L}}\EEQ
 which yields $ 1.1 \ 10^{-11}$ rad $=2.3 \ \mu{\rm as}$ for $M=M_\odot$, explaining the name ``microlensing''
 for solar mass objects and $4.0$ nas, ``nanolensing'', for objects of earth mass.
The $M_\odot$-Einstein radius is $R_E =\theta_Ed_L= 3.05 \cdot 10^{16}$ cm, (Refsdal et al, 2000), 
the microlensing cusp crossing (proper) time is 16.1 year = 5880 days. 

From Eq. (\ref{thetaE=}) we obtain in Table 1 some fiducial values of microlens
masses and microlensing event times. In particular we find that the observed 12 hr
 event detected by Colley and Schild (2003) results from   a nano-lensing
 mass of $2.7 \cdot 10^{-9} M_\odot$. 

\vspace{3mm}

 \begin{center}     
\noindent     
\begin{tabular}{||r|r||}  
 \hline \hline    
mass & duration (days) \\ 
 \hline\hline
$M_\odot$ & 5880  \\ \hline
   0.01$M_\odot$& 588 \\ \hline
    $M_{\rm Jup}$& 182\\ \hline
    0.0001$M_\odot$& 58.8\\ \hline
    $M_{\rm Earth}$& 10.2 \\ \hline
    $10^{-6}$ $M_\odot$& 5.9 \\ \hline
    $10^{-8}$ $M_\odot$& 0.59 \\ \hline
    $2.9 \cdot 10^{-9}$ $M_\odot$&0.2 \\ \hline \hline
\end{tabular}     
\end{center}     
  \centerline{Table 1: event duration (proper time) as function of the lensing mass. }

\myskip{
 \begin{center}     
\noindent     
\begin{tabular}{||r|r||}  
 \hline \hline    
mass $(M_\odot)$ & duration (days) \\ 
 \hline\hline
1 & 10450  \\ \hline
   0.01& 1045 \\ \hline
    $M_{\rm Jup}$& 322 \\ \hline
    0.0001& 105 \\ \hline
    $M_{\rm Earth}$& 18.2 \\ \hline
    $10^{-6}$& 10.4 \\ \hline
    $10^{-8}$& 1.0 \\ \hline
    $2.9 \cdot 10^{-9}$&0.26 \\ \hline \hline
\end{tabular}     
\end{center}     
  \centerline{Table 1: event duration (proper time) as function of the lensing mass. } }

\vspace{6mm}   

The bottom line of Table 1  is for the shortest quasar microlensing event ever observed, 
(Colley and Schild, 2003) with an observed event duration of 12 hours 
and a cosmologically corrected event duration of 5 hours (quasar local clock).
Because it cannot be considered
certain that the observed rapid event is due only to a nano-lensing cusp
crossing, but possibly due at least in part to some effect of orbiting
luminous blobs or obscuring matter (Gould and Miralda-Escud\'e, 1997), the case for Lunar-mass detection is
not certain; however the observed brightness record for the 5 days preceding
the securely detected event allow for the detection of more events of 
similar low brightness amplitude and durations of only hours (Colley and Schild, 2003, Fig. 1).
 
\section{Estimation of the rogue planet mass function}

A statistical microlensing theory for microlensing by masses with Einstein
Ring diameters smaller than the size of the light emissing structures has
been given by Refsdal and Stabell (1991). In its simplest
implementation, it shows the expected average rms brightness fluctuation
amplitude expected for a random distribution of microlenses all assumed to
have the same mass (RS91, Eq. 1). A further elaboration of the theory also
considers the case of microlensing by a distribution of masses having 
a power law distribution function as normally assumed for stars. In the
latter case, the exponent is found to be $\alpha$ = 2.35 and the stellar
mass distribution  is called the Salpeter  function.

Thus if the optical depth is already known, as from a detailed model of the
macro-lensing producing the double quasar image, then a correction to the
measured brightness fluctuation mean amplitude can be determined and from
the RS91 Fig. 2 plot, a correction for the slope $\alpha$ of the mass 
function $F(M)$ determined, if the ratio of the upper and lower mass bounds 
can be estimated.

In our implementation of this scheme, we adopt the following parameters.
We adopt an optical depth of $\sigma=0.707$ (Refsdal et al 2000).
The area of the luminous quasar structure involved in microlensing was
estimated in section 2 to be $3.67\cdot 10^{30} $ cm$^2$ for an equivalent radius
of $1.08\,10^{15}$ cm. It is necessary to calculate this just to ensure
that the ratio of radii is larger than 5 for the RS91 statistical theory to be
applicable, and we find a ratio of 6.7. Thus application of the RS91 theory 
is appropriate. 

We shall need integrals of powers of $M$,

\BEQ
\hspace{-2cm}
I_k=\int_{M_1}^{M_2}\d M\,F(M)\,M^k
=\frac{A}{M_\oplus^{1-\alpha}}\frac{M_2^{k+1-\alpha}-M_1^{k+1-\alpha}}{k+1-\alpha}
=\frac{AM_1^{k+1-\alpha}}{M_\oplus^{1-\alpha}}
\frac{X^{k+1-\alpha}-1}{k+1-\alpha},
\EEQ
with $X=M_2/M_1>1$.  The prefactor $A$ can now be fixed by the normalization $I_0=1$. The average of $M^k$ is

\BEQ
\langle M^k\rangle=\frac{\int_{M_1}^{M_2}\d M\,F(M)\,M^k }{\int_{M_1}^{M_2}\d M\,F(M)}=\frac{I_k}{I_0},
\EEQ
which does not depend on $A$ anyhow. RS91 define the effective mass:

\BEQ
 M_{\rm eff} = \frac{\langle M^2\rangle}{\langle M\rangle}=
 {\overline M} F^2(\alpha)\EEQ
with $ {\overline M}=\sqrt{M_1M_2}$ the geometric average of the upper and lower mass.
 Furthermore,
 \BEQ
 F(\alpha)\equiv
 \left(\frac{I_2}{\bar M I_1}\right)^{1/2}=
 \frac{1}{X^{1/4}} \left[ \frac{2-\alpha}{3-\alpha} \  \frac{X^{3-\alpha}-1}{X^{2-\alpha}-1}\right]^{1/2},
 \EEQ
RS91 point out that $F(\frac{5}{2})=1$ for all $X$. They then argue and support by simulations that for such a mass distribution, 
the predicted rms amplitude of the brightness fluctuations for microlenses of effective mass $M_{\rm eff}$  reads

 \BEA \label{deltam=}
 \delta m_{\rm RS} &=& 2.17 \cdot \sigma^{1/2} \cdot \frac{\theta_E(M_{\rm eff})}{\theta}  \nn\\
& =& 2.17 \cdot \sigma^{1/2} \cdot \frac{\theta_E({\overline M})}{\theta}
  \cdot F(\alpha)  
 \EEA
where $\theta$ is the angular radius of the source. As discussed above, we shall take $\theta=\theta_\eff\equiv R_\eff/d_L=71$ nano-arcsec.

The critical observational result determining the mass function exponent
$\alpha$ is the observed rms amplitude of the observed brightness
fluctuations (the value of $ \delta m_{\rm obs}$ in equation (\ref{deltam=})).
A plot of the rms brightness fluctuations  for the Q0957 gravitational lens system revealed that
a linear relationship exists between the event duration and the amplitude of the measured fluctuations.
This is represented in Fig. 8 of Schild (1999) and the best fit 
curve describing the rms amplitude as a function of event duration
$t_{\rm ev}$ was given as: $ \delta m_{\rm obs} = 0.00133  \,{t_{\rm ev}}\,{\rm day}^{-1} + 0.0060 \,\,{\rm mag}$.
 The measurements extended over a range of $t_{\rm ev}$-values from 2 to 64 days.
   However  it was shown in Schild (1999) that the quasar has fine inner luminous structure presumed to be responsible for the
   fluctuations discussed here, and a more diffusive outer luminous structure.
   However in SLR06 it was shown that 
    the measured quasar brightness fluctuations were
at an observed wavelength of 680 nm which for cosmological redshift 1.43 originated at 280 nm which is at the ``small blue bump'' in the quasar
energy distribution. Moreover the fraction of quasar luminosity originating in the small blue bump is 1/4 of the total,
so a correction factor 4 must be applied for the dilution of the microlensed inner structure by the outer UV-optical continuum measured in reverberation
as described by Schild (2005) and also as modeled by Schild \& Vakulik (2003). 

With correction for the outer quasar luminosity 
the inner microlensed region's rms brightness fluctuation amplitude is (see also the footnote at pages 4--5)

 \BEQ\label{dmobs}
 \delta m_{\rm obs} = 0.00532 \,{t_{\rm ev}}\,{\rm day}^{-1} + 0.0240 \,\,{\rm mag}\EEQ

We have determined our fitting interval with limits $M_1$ and $M_2$ as
follows. The lower mass limit inferred from the hydrodynamical theory for
the formation of small microlensing particles predicts a lower mass limit of $M_1=2.7\,10^{-7}M_\odot$
which corresponds to a wavelet duration of 2 days on observer's clock, the
lowest value for which we have a direct measurement of mean amplitude from
detection of approximately 220 microlensing events in the 4-year intensely
sampled time interval analyzed by Schild (1999). The upper mass limit 
is estimated to be $M_2=2.4 \cdot 10^{-4} M_\odot$ from the maximum of the
wavelets observed. 

For this range of observed masses, the effective mass reads $M_\eff = (M_1 \cdot M_2)^{1/2}=$ $8.05 \cdot10^{-6}M_\odot$ = $2.7 M_\oplus$
and $X=M_2/M_1=890$. For a microlens of this mass
the event duration would be 3.67 days from the calculation of its Einstein
ring diameter and  for the assumed microlens transverse velocity as above.

The event duration that enters (\ref{dmobs}) is defined for each deflector mass as
 
 \BEQ
 t_{\rm ev}
 =(1+z_L)\frac{2 \theta_E(M) d_L}{v}=(1+z_L)\frac{2 \theta_E(\bar M) d_L}{v}\,\sqrt{\frac{M}{\bar M}}
 \EEQ
 with $v=600$ km/s. In our statistical consideration we should average this over the mass distribution (\ref{FMdM}).
 This yields
 
 \BEQ \hspace{-2cm}
 \langle t_{\rm ev}\rangle 
  =
  (1+z_L)\frac{2 \theta_E(\bar M) d_L}{v\sqrt{\bar M}}\frac{I_{1/2}}{I_0}=
  (1+z_L)\frac{2 \theta_E(\bar M) d_L}{vX^{1/4}}G(\alpha)
    = 8.28
  G(\alpha)\,{\rm day} , \EEQ
where
\BEQ
 G(\alpha)=\frac{\alpha-1}{\alpha-\frac{3}{2}} 
 \,\frac{1-X^{3/2-\alpha}}{1-X^{1-\alpha}}
 \EEQ
 We can now equate $\langle\delta m_{\rm obs}\rangle$ to $\delta m_{\rm RS}$, which can be written as
 
 \BEQ\label{H=Hobs}
 H(\alpha) = H_{\rm obs}^{(1)}+H_{\rm obs}^{(2)}\frac{1}{G(\alpha)}
\EEQ
\BEQ
 \hspace{-2cm}  H_{\rm obs}^{(1)}\equiv \frac{0.0049(1+z_L)\, d_L\theta_\eff}{\sqrt{\sigma}v\,{\rm day}}=1.42, 
  \quad H_{\rm obs}^{(2)} =\frac{0.00276\,\theta_\eff X^{1/4}}{\sqrt{\sigma}\theta_E(\bar M)}=0.774
 \EEQ 
  where
 \BEQ
  H(\alpha)\equiv X^{1/4}\frac{F(\alpha)}{G(\alpha)}
 =\frac{\alpha-\frac{3}{2}}{\alpha-1} \,\frac{1-X^{1-\alpha}}{1-X^{3/2-\alpha}}
  \left[ \frac{2-\alpha}{3-\alpha} \  \frac{1-X^{3-\alpha}}{1-X^{2-\alpha}}\right]^{1/2},
  \EEQ
 For our value $X=890$, $H$ has a maximum $H_{\rm max}=6.097$ at $\alpha_c=1.86$, while $H\to 1$  for $\alpha\to\pm\infty$.
 A plot  of both sides of Eq. (\ref{H=Hobs}) is presented in Figure 1. They intersect at $\alpha^\ast=2.98$ and we estimate the error as
 $\alpha^\ast=2.98^{+1.0}{-0.5}$ .
We neglect the intersection for small $\alpha$, so that we have in any case $\alpha>\alpha_c=1.86$ and, in particular, 
we may end up close to the Salpeter value $\alpha=2.35$. However, there is no reason to expect the two values to agree, since they are likely to be
dominated by different processes in star and in primordial rogue planet formation.

\begin{figure}
\centering
\includegraphics[width=7cm]{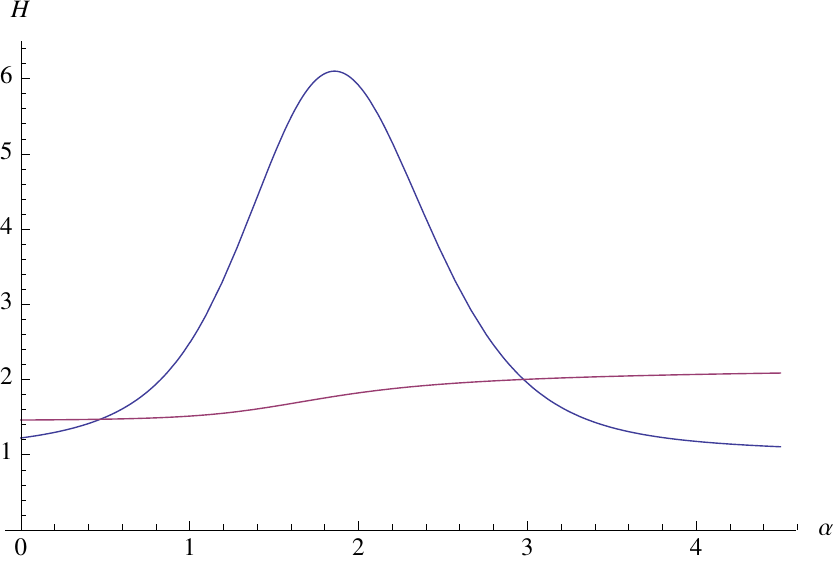}
\caption{Upper curve: $H(\alpha)$ for the case $X=890$. The maximum $6.09697$ occurs at $\alpha=1.85815$; 
the minimal value $H=1$ is attained for $\alpha\to\pm\infty$. Lower curve: the right hand side of Eq. (\ref{H=Hobs}). 
The curves intersect at $\alpha_\ast=2.98$.}
\end{figure}

We list in Table 2 the numerical values determined for the constants
in Eq. (\ref{H=Hobs}). The value for the microlensing optical depth $\sigma$ is the
image A value from Refsdal et al (2000) from the image separation in the overall
gravitational lensing.

\vspace{3mm}

 \begin{center}     
\noindent     
\begin{tabular}{||l|l||}  
 \hline \hline    
Parameter & value \\ 
 \hline\hline
	$M_1 $ & $ 2.7\cdot10^{-7}M_\odot =0.090 \ M_\oplus$\\ \hline
 	$M_2$ &  $ 2.4\cdot10^{-4}M_\odot=80\  M_\oplus $\\  \hline
	${\overline M}=\sqrt{M_1M_2} $ & $ 8.05\cdot10^{-6} M_\odot=2.7 M_\oplus$   \\  \hline
	$X = M_2 / M_1$ & 890 \\  \hline
        $\theta_R$ &  902 nas \\  \hline
     $\theta_E({\overline M}) $ &  6.5 \textrm {nas} \\  \hline
     $\theta_\eff$ &  71.4 nas \\  \hline
	$\sigma$  &  0.707 \\  \hline
	$\alpha_\ast $ & $2.98^{+1.0}_{-0.5}$  \\  \hline
	$\langle t_{\rm ev} \rangle(\alpha_\ast)$ & 11.1 day \\  \hline
	$\langle\delta m_{\rm obs}\rangle(\alpha_\ast) $ & 0.083  \\
	\hline   \hline
\end{tabular}     
\end{center}     
  \centerline{Table 2: 	Empirical values of microlensing parameters}
    
\vspace{6mm}

Our power law slope $\alpha_\ast=2.98$ is steeper than the Salpeter value of 2.35. There is no reason
to expect the two values to agree, since they are likely to be dominated by
different processes in star and in primordial rogue planet formation.

\section{Summary and Conclusions}

With the quasar microlensing detection of a cosmologically significant
fraction of the missing baryons (baryonic dark matter) now confirmed
by high-cadence MACHO searches to the galactic center and to the LMC 
(Sumi et al, 2010, 2011), we investigate methods to permit
investigation of the mass distribution function of the rogue planet
population. The only theory that has predicted that the baryonic dark
matter should exist as a population of rogue planets  (Gibson, 1996) predicts that at time
of formation their mass was $10^{-7}$ M$_\odot$, and that they should be
found primarily in primordial Jeans clusters of approximately $6\cdot 10^5 M_\odot$
(Nieuwenhuizen, Schild and Gibson, 2011).
Such Jeans clusters have also been found in quasar milli-lensing at significant
optical depth (Mao and Schneider, 1998). In the high-temperature and
high-density universe at $z \sim 1000$, these sticky hydrogen spheres were
formed by the usual void-condensation separation process and would have
immediately interacted to form pairs, triples and pairs-of-pairs to eventually
accumulate in an accretional cascade that quickly produced the first stars
and left behind rogue planets of larger mass.
Most Jeans clusters should have remained dark, but rapidly form stars when
disturbed. A prediction of this theory is that all or most stars should be binaries.

We also demonstrate how the expected microlensing signal apparent in
observed brightness fluctuations in quasars can be analyzed with a statistical theory
devised by RS91. We find that the observed amplitudes of the microlensing
fluctuations are approximately a factor 2 smaller than predicted, and that
this mimics the amplitudes expected if the particles are not all of the
same mass, but instead are distributed according to a mass function with
slope approximately $2.98^{+1.0}_{-0.5}$, where the Salpeter slope is 2.35.   

Our study thus far has been applied to the Q0957+561 A,B quasar system (the
First Lens) for which a great deal of information on the luminous quasar
structure has been inferred in SLR06. However this radio source quasar in
the Lo-Hard spectral state is probably not optimum since its luminous
inner accretion disc edge is probably finer than the High Soft radio quiet
objects, which are also more common. Thus in the future we will extend our study to the
several radio-quiet lensed quasars with time delay measurements and
significant observed microlensing residuals.

\vspace{1cm}

\newcommand{\asas}{Astron. \& Astrophys}
\newcommand{\apj}{Astroph. J.}
\newcommand{\pasp}{pasp}

\end{document}